\documentclass[aps,prd,twocolumn,showpacs,amsmath,%
amssymb,nofootinbib,floatfix]{revtex4}

\usepackage[T2A]{fontenc}
\usepackage[cp1251]{inputenc}
\usepackage[english]{babel}
\usepackage{amsbsy}
\usepackage{graphicx}
\usepackage{calrsfs}
\usepackage{natbib}
\usepackage[hypertex,breaklinks]{hyperref}

\begin{document}

\def\b{\boldsymbol}
\def\d{\partial}
\def\p{\varpi}
\def\e{\varepsilon}
\def\k{\kappa}
\def\ds{\displaystyle}
\def\t{\tilde}
\def\apjs{ApJS}

\title{Mechanics and kinetics in the Friedmann--Lema\^\i tre--Robertson--Walker
space-times}

\author{S.R.~Kelner}
\affiliation
{Max-Planck-Institut f\"ur Kernphysik,
Saupfercheckweg 1, D-69117 Heidelberg, Germany }
\email{Stanislav.Kelner@mpi-hd.mpg.de}

\author{A.Yu.~Prosekin}
\affiliation{Max-Planck-Institut f\"ur Kernphysik,
Saupfercheckweg 1, D-69117 Heidelberg, Germany}
\email{Anton.Prosekin@mpi-hd.mpg.de}

\author{F.A.~Aharonian}
\affiliation{Dublin Institute for Advanced Studies, 31 Fitzwilliam Place,
Dublin 2, Ireland}
\altaffiliation
{Max-Planck-Institut f\"ur Kernphysik,
Saupfercheckweg 1, D-69117 Heidelberg, Germany}
\email{Felix.Aharonian@mpi-hd.mpg.de}

\date{\today}

\begin{abstract}
Using the standard canonical formalism, the equations of mechanics and kinetics
in the Friedmann--Lema\^\i tre--Robertson--Walker (FLRW) space-times
in Cartesian coordinates have been obtained.
The transformation law of the generalized momentum under the shift of the
origin of the coordinate system has been found, and the form invariance of
the  Hamiltonian function relative to the shift transformation has been proved.
The general solution of the collisionless Boltzmann equation has been found.
In the case of the homogeneous distribution the solutions of the kinetic
equation for several simple, but important for applications, cases have been
obtained.
\end{abstract}


\maketitle

%\section{Introduction}
\section{\label{Intro}Introduction}

The covariant general relativistic Boltzmann equation for the one-particle
distribution function has been found in  Ref.~\cite{Walker}. Eq.~(16) from
Ref. \cite{Walker} reads
\begin{equation}\label{W1}
 \frac{\d\chi}{\d x^i}\,p^i-\frac{\d\chi}{\d p^i}\,\Gamma^i_{jk}p^j p^k=0\,.
\end{equation}
However, the physical meaning of the function $\chi$ and correct interpretation
of the equation had been revealed  much later (see Refs.~\cite{Debbasch1,Debbasch2}
and references cited
there, as well as  Ref.~\cite{Dodin}, where  a critical review of  previous
studies conducted for the basic plasma modes in the expanding
Universe is given). Throughout the present paper, except for Appendix~\ref{ApA},
we consider the motion of particles and the kinetics in the FLRW space-times
in Cartesian coordinates. The uniqueness of the FLRW metrics \cite{Collins},
its  homogeneity and isotropy are more clearly exposed in these  coordinates,
therefore we do not use generally covariant notations.

Our approach is based on the standard scheme of the classical mechanics:
the generalized coordinates and Lagrangian function $\to$ the generalized
momentum $\to$ the Hamiltonian function $\to$ the phase space. Thus there
is no problem to derive the collisionless Boltzmann equation and to interpret
the distribution function. Moreover, the Hamilton-Jacobi equation allows us,
as in the case of the conventional space, to find in the explicit form all
six integrals of motion and thereby to obtain the general solution of the
Boltzmann equation. The solution is particularly simple in the spherically
symmetric case. 

It is well known that the metrics FLRW is form-invariant relative to the
shift of the origin of the coordinate system, and it is known how the Cartesian
coordinates are transformed under the shift \cite{Weinberg}. The mechanics
also appears to be form-invariant. This circumstance is always implicitly assumed, 
however, to our knowledge,  the direct prove of the invariance has not been
demonstrated.  In this paper we  prove the form-invariance of the 
Hamiltonian function and find the  law of the momentum transformation under the 
shift of the origin of coordinates. If
one moves  the reference point to  the point where the observer is located,
we can easily  interpret the results of calculations since in the vicinity 
of the observation point the space can be considered as Euclidean one.  

In the curved space, if the distribution of  particles is homogenous, 
it is isotropic as well (see Appendix~\ref{ApB}).
In this case,  the derivation of the  collisionless Boltzmann equation 
is trivial, and the collision integrals can be  described  in the same 
way as in the flat space.  It is impossible to find  analytical 
solutions to the Boltzmann equation with
collision integrals, therefore we restricted ourselves by several simple,
but important for applications, cases when the solution can be obtained by
quadratures.  

For convenience,  some of the principal calculations are presented in four Appendices.
In Appendix~\ref{ApA}  we give a simple derivation of the Boltzmann equation for
space with an arbitrary metric written in arbitrary coordinates.  The Hamilton
function has been obtained, and it has been shown
that the equation of motion and the Boltzmann equation can be written in
the standard form.  Finally, in Appendix~\ref{ApD} we note  the surprising effect 
 that a  photon, emitted  by the source with superluminal recession velocity
 in the direction to the observer, during a certain time interval moves away from the observer.

\section{Mechanics of free particles in FLRW space-times}
In Cartesian coordinates the Friedmann--Lema\^\i tre--Robertson--Walker 
metric can be represented in the form

\begin{equation}\label{pm1}
 ds^2=dt^2-a^2(t)\left(d\b r^2+\k\frac{(\b rd\b r)^2}{1-\k\b r^2}\right),
\end{equation}
where $\k$ is the discrete quantity which describes possible isotropic models
and has the following values: $\k=1$ for the closed model (positive curvature),
$\k=-1$ for the open model (negative curvature), and $\k=0$ for the flat
space. The function $a(t)$, which is determined by Friedmann equation, is
assumed to be given.  We use  the three-dimensional vector notations, 
$ \b r=(x,y,z)\,,\; (\b rd\b r)=x\,dx+y\,dy+z\,dz$, etc., where $\b r$
is considered to be vector in the sense that under rotations
relative to the origin of coordinates the components of $\b r$ are transformed
the same way as  the vector components in Euclidean space.

The action functional of a particle in the gravitational field
is given by (assuming $c=1$)
\begin{equation}
 S=-m\int\! ds=\int\! L\,dt\,,
\end{equation}
where Lagrangian function is
\begin{equation}\label{pm2}
 L=-m\,\sqrt{1-a^2(t)\left(\b v^2+\k\frac{(\b r\b v)^2}
 {1-\k\b r^2}\right)}\,.
\end{equation}
Considering $\b r$ and $\b v=d\b r/dt$ as generalized coordinates and velocities,
we can make use of the formalism of the classical mechanics (see, e.g., \cite{Landau1}, \cite{Goldstein}).
Then the generalized momentum is expressed as

\begin{equation}\label{pm3}
\b p=\frac{\d L}{\d\b v}=\frac{\ds ma^2\left(\b v+\k\frac{\b r(\b r\b v)}
{1-\k\b r^2}\right)} {\sqrt{\ds 1-a^2(t)\left(\b v^2+\k\frac{(\b r\b v)^2}
{1-\k\b r^2}\right)}}\,,
\end{equation}
and the energy is
\begin{equation}\label{pm4}
E=\b v\,\frac{\d L}{\d\b v}-L=\frac{m}{\sqrt{\ds 1-a^2(t)
\left(\b v^2+\k\frac{(\b r\b v)^2}{1-\k\b r^2}\right)}}\,.
\end{equation}
Note that $E\ge m$, as it is in Minkowski space. It is easy to ascertain
by direct check that  ($a\equiv a(t)$)
\begin{equation}\label{zz1}
 E^2-\frac1{a^2}\,(\b p^2-\k(\b p\b r)^2)=m^2\,,
\end{equation}
therefore  the Hamiltonian function  has the following form
\begin{equation}\label{pm5}
{\cal H}(\b p,\b r,t)=\frac1a\sqrt{\b p^2-\k(\b p\b r)^2+m^2a^2}\,.
\end{equation}
The dependence of generalized coordinates and momenta on time is found from
canonical equations of motion  
\begin{equation}\label{r1}
 \dot{\b r}=\frac{\d\cal H}{\d\b p}=\frac1a\,\frac{\b p -\k\b r\,(\b p\b r)}
 {\sqrt{\b p^2-\k(\b p\b r)^2+m^2a^2}}\,,
\end{equation}
\begin{equation}\label{r2}
 \dot{\b p}=-\frac{\d\cal H}{\d\b r}=\frac1a\,\frac{\k\b p\,(\b p\b r)}
 {\sqrt{\b p^2-\k(\b p\b r)^2+m^2a^2}}\,,
\end{equation}
which admit an exact analytical solution in general case (see below).
For massless particles the Hamiltonian function  becomes
\begin{equation}\label{pm5a}
{\cal H}=\frac1a\sqrt{\b p^2-\k(\b p\b r)^2}\,.
\end{equation}
This formula can also be applied to ultrarelativistic particles.

Since the Hamiltonian function explicitly depends on time, the energy of particle
is not conserved. To determine the time dependence of the energy, let us
consider the differential equation which at any $\k$ has the form

\begin{equation}\label{pm5b}
\frac{dE}{dt}=\frac{\d{\cal H}}{\d t}=-\frac{\dot
a}{a}\,\left(E-\frac{m^2}{E}\right).
\end{equation}
The solution of this equation gives the relation between values of the energy
of a freely moving particle at different moments of time:
\begin{equation}\label{pm5c}
E(t)=\left[\left(\frac{a(t')}{a(t)}\right)^{\!2} (E^2(t')-m^2)+m^2
\right]^{\!1/2}.
\end{equation}
For photons (or ultrarelativistic particles) this leads to the well known
relation between the energy and the scale factor (redshift):

\begin{equation}\label{pm5d}
a(t)\,E(t)={\rm const}\,.
\end{equation}
Often  this relation is interpreted as a consequence of the Doppler effect. However
this  seems to us not correct since Eq.~(\ref{pm5c}) for $m\ne 0$ cannot
be obtained using Lorentz transformation. In the nonrelativistic case, denoting
$E=m+E_{\rm kin}$ and assuming that $E_{\rm kin}\ll m$, we get
\begin{equation}\label{pm5non}
a^2(t)\,E_{\rm kin}(t)={\rm const}\,,
\end{equation}
which means that for nonrelativistic particles the decrease of kinetic
energy with time is faster. 

\section{Form-invariance of mechanics}
The space with metric given by Eq.~(\ref{pm1}) is homogeneous and isotropic.
The isotropy of the space is obvious  since the quantities
$d\b r^2$, $\b r^2$ and $(\b r\,d\b r)$ in Eq.~(\ref{pm1}) do not change under
rotation relative to the origin of coordinates. The homogeneity implies  that  the origin of 
coordinates can be chosen at any point of the space, and the metric would have 
the same form of Eq.~(\ref{pm1}). The proof of homogeneity can be found 
in Ref.~\cite{Weinberg}. If the origin of coordinates is shifted from the
point $\b r=0$ to the point $\b r=\b b$, then the new coordinates $\b r'$
are expressed through the old coordinates in the following way \cite{Weinberg}:
\begin{equation}\label{pm6}
\b r'=\t{\b r}(\b r,\b b) \equiv \b r-\b b\left(\sqrt{1-\k r^2}+
\frac{\k(\b b\b r)} {\sqrt{1-\k b^2}+1} \right).
\end{equation}
If $\b r=\b b$, we have $\b r'=0$. The inverse transformation
\begin{equation}\label{pmi6}
\b r=\t{\b r}(\b r',-\b b)
%=\b r'+\b b\left(\sqrt{1-\k r'^2}-\frac{\k(\b b\b r')}
%{\sqrt{1-\k b^2}+1}\right).
\end{equation}
is obtained from Eq.~(\ref{pm6}) by replacing $\b b\to -\b b$. The ``volume''
elements in the new and old coordinates are connected by relations
\begin{gather}\label{dr1dr}
 d^3r'=\left(\sqrt{1-\k b^2}+\frac{\k(\b b\b r)}{\sqrt{1-\k r^2}}\right) d^3r\,,
\end{gather}
\begin{gather}\label{drdr1}
d^3r=\left(\sqrt{1-\k b^2}-\frac{\k(\b b\b r')}{\sqrt{1-\k r'^2}}\right)
d^3r'
\end{gather}
(note that Eqs.~(\ref{dr1dr}) and (\ref{drdr1}) are equivalent).

Let us find the transformation laws  for the  velocity and momentum under
shifts.
The velocity is transformed as contravariant vector. Assuming
$\b r=\b r(t)$, $\b r'=\b r'(t)$ in Eq.~(\ref{pm6}) and differentiating it
with respect to $t$, we find
\begin{equation}\label{pm6a}
\b v'=\t{\b v}(\b v,\b r,\b b) \equiv \b v+\k\b b\left(
\frac{(\b r\b v)}{\sqrt{1-\k r^2}} -\frac{(\b b\b v)}
{\sqrt{1-\k b^2}+1}\right).
\end{equation}
The momentum, as it follows from  Eq.~(\ref{pm3}), is a covariant
vector, therefore the transformation law of $\b p$ is
\begin{equation}\label{zz4}
p'_\alpha={(M^{-1})_\alpha}^\beta\,p_\beta\,,
\end{equation}
where matrix $M^{-1}$ is inverse to ${M_\alpha}^\beta=\d v'_\beta/\d
v_\alpha^{}$. Eq.~(\ref{zz4}) can be written in the  explicit form : 
 \begin{multline}\label{pm7}
 \b p'=\t{\b p}(\b p,\b r,\b b) \equiv\b p-\frac{\k(\b b\b p)}
 {\sqrt{(1-\k b^2)(1-\k r^2)}+\k(\b b\b r)}\\
 \times\left(\b r-\frac{\b b\,\sqrt{1-\k r^2}}{\sqrt{1-\k b^2}+1}\right).
\end{multline}
As in the case of the transformation  Eq.~(\ref{pm6}), the inverse transformations
of Eqs.~(\ref{pm6a}) and (\ref{pm7}) can be obtained by replacement of $\b
b$ with $-\b b$ and interchange of primed and unprimed quantities:
\begin{equation}\label{pvinv}
 \b v= \t{\b v}(\b v',\b r',-\b b)\,,\quad \b p= \t{\b p}(\b p',\b r',-\b b)\,.
\end{equation}
From Eqs.~(\ref{pm7}) and  (\ref{pvinv}) the following 
relations between the volume elements
in the momentum space can be found:
\begin{gather}\label{dp1dp}
 d^3p'=\frac{\sqrt{1-\k r^2}}{\sqrt{(1-\k r^2)(1-\k b^2)}+\k(\b b\b r)}
\,d^3p\,,
\end{gather}
\begin{gather}\label{dpdp1}
 d^3p=\frac{\sqrt{1-\k r'^2}}{\sqrt{(1-\k r'^2)(1-\k b^2)}-\k(\b b\b r')}
\,d^3p'\,.
\end{gather}

The change of variables from $(\b p,\b r)$ to $(\b p',\b r')$ is a canonical
transformation that can be proved by direct computation of Poisson brackets.
However, it can be demonstrated much easier by noting that the transformation
can be carried out by the generating function
\begin{equation}\label{pm9}
{\cal S}(\b p,\b r')=(\b p\b r')+(\b p\b b)\left(\sqrt{1-\k r'^2}-
\frac{\k(\b b\b r')}{\sqrt{1-\k b^2}+1}\right).
\end{equation}
The equations
\begin{equation}\label{pm10}
\b r=\frac{\d{\cal S}}{\d\b p}\,,\qquad \b p'=\frac{\d{\cal S}}{\d\b r'}
\end{equation}
following from this are equivalent to Eqs.~(\ref{pm6}) and (\ref{pm7}).

Since ${\cal S}$ does not depend on time, the old and new Hamiltonian functions
are equal: ${\cal H}'={\cal H}$. By the direct check one can ascertain the validity
of the equation 
\begin{equation}\label{pm8}
\b p'^2-\k(\b p'\b r')^2=\b p^2-\k(\b p\b r)^2\,.
\end{equation}
Therefore, in the new reference system
\begin{equation}\label{zz3}
 {\cal H}'(\b p',\b r',t)\!=\!{\cal H}(\b p',\b r',t)\!=\!
 \frac1a\sqrt{\b p'^2-\k(\b p'\b r')^2+m^2a^2}\,,
\end{equation}
i.e. the Hamiltonian function  is form-invariant. Thus we arrive at the natural conclusion
that not  only the geometry but also the mechanics  in FLRW
space--times does not change under the  shift of the origin of coordinates. It should be kept
in mind that in the curved space not only generalized coordinates but also
generalized momenta depends on choice of origin of coordinates.

The transformations given by Eqs.~(\ref{pm6}), (\ref{pm6a}) and (\ref{pm7}) are convenient
for analysis of the results obtained in the curved space. 
Let us assume that we know the solution of a problem in the reference frame
with the origin located at the source, and an observer is located at the
point $\b r$. For analysis of the result it is convenient to move to another
coordinate system shifting the origin to the location of observer, i.e. 
to assume $\b b=\b r$. The space in the small neighborhood of $\b r'=0$ of
new reference frame can be considered as Euclidean one that appreciably simplifies
the analysis. 

The replacement of $\b b=\b r$   in Eqs.~(\ref{pm6a}) and (\ref{pm7}) gives the
generalized velocity and momentum in the observation point:
\begin{gather}\label{vb}
 \b u = \t{\b v}(\b v,\b r,\b r)=
 \b v+\frac{\b r(\b r\b v)}{r^2}\left(\frac{1}{\sqrt{1-\k r^2}}-1\right),
\end{gather}
\begin{gather}\label{pb}
\b q = \t{\b p}(\b p,\b r,\b r)=
\b p+\frac{\b r(\b r\b p)}{r^2}\left(\sqrt{1-\k r^2}-1\right).
\end{gather}
The vectors $\b u$ and $\b q$ are parallel,  while their squares  are
\begin{equation}\label{qq}
\b u^2=\b v^2+\frac{\k(\b v\b r)^2}{1-\k r^2}\,,\quad
\b q^2=\b p^2-\k(\b p\b r)^2\,.
\end{equation}
It should be noted that quantities $\b u^2$ и $\b q^2$ are invariants relative
to shift. Multiplying Eq.~(\ref{pb}) by vector $\b r$, we find
\begin{equation}\label{pb1}
 (\b q\b r)=(\b p\b r)\sqrt{1-\k r^2}\,.
\end{equation}
Rewriting $(\b p\b r)=pr\cos\theta$, $(\b q\b r)=qr\cos\theta'$ and taking
into account that $q=p\sqrt{1-\k r^2\cos^2\theta}$ in such notation, we obtain
the following relation between the angles in the new and old coordinate systems:
\begin{equation}\label{pb2}
 \cos\theta'=\cos\theta\,\sqrt{\frac{1-\k r^2}{1-\k r^2\cos^2\theta}}\,,
\end{equation}
and
\begin{gather}\label{pb3}
 \cos\theta=\frac{\cos\theta'}{\sqrt{1-\k r^2\sin^2\theta'}}\,,
\end{gather}
\begin{gather}\label{pb4}
\sin\theta=\sin\theta'\,\sqrt{\frac{1-\k r^2}{1-\k r^2\sin^2\theta'}}\,.
\end{gather}

Let us denote by $\b V$ and $\b P$ the usual (not generalized) velocity and
momentum of the particle  registered by the observer. Then 
\begin{equation}\label{usual}
 \b V=a\,\b u\,,\quad \b P=\b q/a\, .
\end{equation}
Note that $(\b P\b V)=(\b q\b u)=(\b p\b v)$. From Eqs.~(\ref{vb})
and (\ref{pb}) one can find the generalized velocity and momentum at the point
$\b r$ expressed in terms of $\b V$ and $\b P$:
\begin{gather}\label{vb1}
\b v=\frac1a\left[\b V+\frac{\b r(\b r\b V)}{r^2}\left(\sqrt{1-\k r^2}
-1\right)\right],
\end{gather}
\begin{gather}\label{pb4z}
\b p=a\left[\b P+\frac{\b r(\b r\b P)}{r^2}\left(\frac{1}
{\sqrt{1-\k r^2}}-1\right)\right].
\end{gather}
The relations between quantities $\b V$, $\b P$ and $E$ are the same as in
special relativity:
\begin{equation}\label{EVP}
 E=\frac{m}{\sqrt{1-V^2}}=\sqrt{P^2+m^2}\,,\quad
 \b V=\frac{\b P}{E}\,.
\end{equation}

As an example, let us consider the solution given by  Eqs.~(\ref{r1}), (\ref{r2}) in
the case of negative curvature ($\k=-1$). The origin of coordinates
is taken at the position of the particle at the initial moment of time  $t_i$,
i.e. $\b r(t_i)=0$. Then the motion is radial and we seek the solution in
the following form 
\begin{equation}\label{r3}
 \b r(t)=\b n\rho(t)\,,\qquad \b p(t)=\b n\p(t)
\end{equation}
with initial condition $\rho\big|_{t=t_i}=0$, $\p\big|_{t=t_i}=\p_0^{}$,
where $\b n$ is an arbitrary unit vector, $\p_0^{}$ is an arbitrary constant.
The differential equations for $\rho$ and $\p$ are given by
\begin{gather}\label{r4}
\frac{d\rho}{dt}=\frac1a\,\frac{\p\,(1+\rho^2)}{\sqrt{\p^2(1+\rho^2)+m^2a^2}}\,,
\end{gather}
\begin{gather}\label{r4a}
 \frac{d\p}{dt}=-\frac1a\,\frac{\p^2\rho}{\sqrt{\p^2(1+\rho^2)+m^2a^2}}\,.
\end{gather}
Dividing  one equation to another, we find
\begin{equation}\label{r5}
 \frac{d\p}{d\rho}=-\frac{\p\rho}{1+\rho^2}\,,
\end{equation}
from where it follows that $\p^2(1+\rho^2)=\p_0^2={\rm const}$.

It is convenient to introduce a new function
$\eta$ defined as
\begin{equation}\label{eta}
 \eta(t,t_i)=\int_{t_i}^t\!\frac{dt'}{a(t')\sqrt{1+(ma(t')/\p_0^{})^2}}\,.
\end{equation}
Then the functions  from  Eq.~(\ref{r3}) can be expressed through $\eta$:
\begin{equation}\label{r6}
\rho(t)=\sinh\eta\,,\qquad \p(t)=\p_0^{}/\cosh\eta\,.
\end{equation}
The generalized velocity is
\begin{equation}\label{r6a}
 \b v=\frac{d\b r}{d\eta}\,\frac{d\eta}{dt}=
 \frac{\b n\cosh\eta}{a\sqrt{1+(ma/\p_0^{})^2}}\,.
\end{equation}
Substituting the solution into Eqs.~(\ref{vb}) and (\ref{pb}) and assuming $\k=-1$,
we find the velocity and momentum at an arbitrary moment of time 
\begin{equation}\label{vp1}
 \b P=\frac{\b n\p_0^{}}{a}\,,\quad
 \b V=\frac{\b n}{\sqrt{1+(ma/\p_0^{})^2}}\,.
\end{equation}
If at the initial moment of time the particle is at the point $\b r=\b b$,
the solutions have  the form
\begin{gather}\label{rb1}
 \b r(t)=\t{\b r}(\b n\sinh\eta,-\b b)\,,
\end{gather}
\begin{gather}\label{rb2}
 \b p(t)=\t{\b p}(\b n\p_0^{}/\cosh\eta,\b n\sinh\eta,-\b b)\,.
\end{gather}

This expressions  are obtained from Eq.~(\ref{r3}) by shifting the origin of coordinates 
to $-\b b$. They describe the general solution of Hamilton equations which
depends on six arbitrary constants: three components of vector $\b b$, two
angles defining the direction of $\b n$, and $\p_0^{}$.

For the space with positive curvature ($\k=+1$) the calculations are similar.
Eq.~(\ref{vp1}) remains correct in this case, but instead of Eq.~(\ref{r6}) we
have
\begin{equation}\label{r6b}
\rho=\sin\eta\,,\qquad \p=\p_0^{}/\cos\eta\,,
\end{equation}
where $\eta$ is defined as before by Eq.~(\ref{eta}). 

\section{Distribution function}
Let $f(\b p,\b r,t)$ is a distribution function of particles in the phase
space. By definition, the quantity 
\begin{equation}\label{pm13}
dN=f(\b p,\b r,t)\,d^3p\,d^3r
\end{equation}
implies the number of particles found at the moment $t$ in the volume element $d^3p\,d^3r$
of the phase space. The Jacobian of the canonical transformation equals unity,
therefore the phase volume does not change under shift, i.e.
\begin{equation}\label{vol}
 d^3p\,d^3r=d^3p'\,d^3r'\,,
\end{equation}
where the canonical variables  $(\b p',\b r')$ are connected to  $(\b p,\b r)$
through  Eqs.~(\ref{pm6}) and  (\ref{pm7}). One can directly verify Eq.~(\ref{vol})
by multiplying  Eqs.~(\ref{dr1dr}) and (\ref{dp1dp}), or (\ref{drdr1}) and (\ref{dpdp1}).
Since $dN$ and $d^3p\,d^3r$ are invariants, the distribution function is
also invariant relative to the shift of the origin of coordinates:
\begin{equation}\label{pm14}
f(\b p,\b r,t)=f'(\b p',\b r',t)\,.
\end{equation}

The Boltzmann equation for $f$ has  the following standard form 
(see~Appendix~\ref{ApA})
\begin{equation}\label{pm14a}
\hat Lf\equiv
\left(\frac{\d}{\d t}+\frac{\d{\cal H}}{\d\b p}\,\frac{\d}{\d\b r}-
\frac{\d{\cal H}}{\d\b r}\,\frac{\d}{\d\b p}\right)\!f=0\,.
\end{equation}
This equation describes the evolution of the distribution function of free-moving
particles\footnote{In the absence of collisions this equation is equivalent
to the Vlasov equation or Liouville equation for one particle.}. To take
into
account interactions, the  collision integral should be added to the right part.

For a single particle  moving according to Eq.~(\ref{r3}), the distribution
function is
\begin{equation}\label{f1}
 f(\b p,\b r,t)=\delta(\b p-\b n \p_0^{}/\cosh\eta)\,
 \delta(\b r-\b n \sinh\eta)\,.
\end{equation}
If at the initial moment of time a particle is found at the point $\b r=\b
b$ and has  momentum $\b P_0=\b n\p_0/a(t_i)$, then the distribution function
is
\begin{multline}\label{f2}
 f(\b p,\b r,t)=\delta\big(\t{\b p}(\b p,\b r,\b b)-\b n
\p_0^{}/\cosh\eta\big)\,\\
 \times\delta\big(\t{\b r}(\b r,\b b)-\b n \sinh\eta\big)\,,
\end{multline}
where $\t{\b r}$ and $\t{\b p}$ is defined in Eqs.~(\ref{pm6}) and (\ref{pm7}).
Here the invariance of $f$ relative to the shift has been used. Eqs.~(\ref{pmi6}) and 
(\ref{pvinv}), as well as  the fact that Jacobian is unity,  allow us to write $f$ in the
form:
\begin{multline}\label{f3}
 f(\b p,\b r,t)=
 \delta\big(\b p-\t{\b p}(\b n\p_0/\cosh\eta,\b n\sinh\eta ,-\b b)\big)\,\\
 \times\delta\big(\b r-\t{\b r}(\b n\sinh\eta,-\b b)\big)\,.
\end{multline}

The general solution of the Eq.~(\ref{pm14a}) can be found in the case of
free-moving particles. To find the characteristics of the equation, let us
return to the problem considered in the previous section and solve it by
the use of Hamilton-Jacobi equation, which for the Hamiltonian function 
of Eq.~(\ref{pm5}) has
the form (for sake of definiteness we restrict our considerations to the
case of $\k=-1$)
\begin{equation}\label{hj1}
 \frac{\d S}{\d t}+\left[\frac{1}{a^2(t)} \left((\nabla S)^2+(\b r\nabla S)^2
\right) +m^2 \right]^{\!1/2}= \,0.
\end{equation}
As usual, for integrable systems the complete integral of the equation can be 
found by separation of variables \cite{Landau1}.
Solving Eq.~(\ref{hj1}), we find
\begin{equation}\label{hj2}
 S(\b s,\b r,t)=s\,{\rm arсsinh}(\b\nu\b r)
-\int^t\!\sqrt{s^2/a^2(t')+m^2}\,dt'\,.
\end{equation}
Here $s$ is an arbitrary constant,  $\b\nu$ is an arbitrary
unit vector. Without loss of generality it can be assumed that
$s\ge0$. It is convenient to consider the expression in the right part of
Eq.~(\ref{hj2}) as a function of $t$ and the vectors $\b s=s\b\nu$ and $\b
r$. The lower limit of integration over $dt'$ is taken for convenience. It is
easy to show directly that Eq.~(\ref{hj2}) satisfies to Eq.~(\ref{hj1}).

The momentum is
\begin{equation}\label{hj3}
 \b p=\frac{\d S}{\d\b r}=\frac{\b s}{\sqrt{1+(\b\nu\b r)^2}}\,.
\end{equation}
From this it follows that  $\b\nu=\b s/s=\b p/p$,  thus 
\begin{equation}\label{hj4}
 \b s=\b p\sqrt{1+(\b\nu\b r)^2}\,.
\end{equation}
To determine the particle motion, let us differentiate $S$ with respect to
arbitrary constants and equate the  result to another constants:
\begin{equation}\label{hj5}
\frac{\d S}{\d\b s}=\b\nu\,{\rm arсsinh}(\b\nu\b r)+ \frac{\b r-\b\nu(\b\nu\b
r)}
{\sqrt{1+(\b\nu\b r)^2}}-\eta\b\nu=\b\xi\,,
\end{equation}
where
\begin{equation}\label{hj6}
 \eta=\int^t\!\frac{dt'}{a(t')\,\sqrt{1+(ma(t')/s)^2}} \,.
\end{equation}
The six quantities $\b s$ and $\b\xi$ are integrals of motion of the problem
and at the same time they are characteristics of Eq.~(\ref{pm14a}). Therefore
the general solution of the Boltzmann equation can be presented in the form
\begin{equation}\label{hj7}
 f(\b p,\b r,t)=\Phi(\b s,\b\xi)\,,
\end{equation}
where $\Phi$ is an arbitrary function\footnote{The general solution based
on Killing vector constants of the motion is obtained as well in Ref.~\cite{Gen_Sol}.
However,  the solution is expressed as a function of arguments which are not canonically
conjugated quantities. Therefore one needs  an additional analysis to
find the relationship between the solution and the distribution function.}. Here it is implied that $\b s$ and
$\b\xi$ are expressed via generalized coordinates and momenta via
Eqs.~(\ref{hj4}) and (\ref{hj5}).
 
Assuming $m=0$, the solution given by Eq.~(\ref{hj7}) can be applied to study the evolution
of distribution function of photons and neutrinos in FLRW space-time in the
case of inhomogeneous and anisotropic distribution.

%%%%%%%%%%%%%%%%%%%%%%%%%%%%%%%%%%%%%%%
For the collision integrals it  is convenient to use, instead
of generalized momentum,  the energy $E$
and unit vector $\b n=\b P/P=\b q/q$ in the direction of momentum.  Let us introduce the distribution function $F$ according
to the relation
\begin{equation}\label{pm20}
dN=F(E,\b n,\b r,t)\,dE\,\frac{d\Omega}{4\pi}\,dV\,.
\end{equation}
$dN$  is  the number of particle at the moment $t$ located in the volume
$dV$ with the energy restricted in the interval $dE$ and the direction of
the momentum $\b P$ enclosed in the solid angle $d\Omega$. It is follows
from the metric of Eq.~(\ref{pm1}) that the volume element is 
\begin{equation}\label{q1}
dV=\frac{a^3(t)\,d^3r}{\sqrt{1-\k r^2}}\,.
\end{equation}
Let us denote $dV_a=dV/a^3$ which is dimentionless volume element (in the
units of $a^3$). It is easy to find from Eq.~(\ref{pb}) that
\begin{equation}\label{q2}
d^3q=\sqrt{1-\k r^2}\,d^3p\,,
\end{equation}
therefore the volume element of the phase space is
\begin{equation}%\label{q3}
d^3p\,d^3r=d^3q\,dV_a=d^3P\,dV\,.
\end{equation}
Using this equation  and definitions given by  Eqs.~(\ref{pm13}) and  (\ref{pm20}), 
we find the relation between  functions $f$ and $F$: 
\begin{equation}\label{pm23}
f(\b p,\b r,t)=F(E,\b n,\b r,t)/(4\pi PE)\,.
\end{equation}

Since $E$ and $|\b P|$ do not change under shifts, $F$ is an invariant
as is $f$
\begin{equation}%\label{q3a}
 F(E,\b n,\b r,t)=F'(E,\b n',\b r',t)\,,
\end{equation}
where $\b n'$ is unity vector in the direction of
$\b q' = \t{\b p}(\b p',\b r',\b r')$. Further in this section we use both
distribution functions. 

To derive the collision integral in the FLRW space-time it is convenient to
shift the origin of the reference frame to the collision point. Then the collision integral
can be written as in the flat space. After that one should move to the initial
reference frame.

%%%%%%%%%%%%%%%%%%%%%%%%%%%%%%%%%%%%%%%

\subsection{Homogeneous distribution\label{uniform}}

Let us consider the case of the homogeneous space distribution
of particles. In this case the distribution function should be form-invariant
relative to shift, i.e. it has an identical form in different reference frames.
Moreover, along with Eq.~(\ref{pm14}) the stronger condition should be
fulfilled:
\begin{equation}\label{pm15}
f(\b p,\b r,t)=f(\b p',\b r',t)\,.
\end{equation}
If $\b p'$ and $\b r'$ are expressed in accordance with Eqs.~(\ref{pm6}) and  
(\ref{pm7}) in terms of $\b p$, $\b r$ and $\b b$, then the right part should not depend
on  $\b b$. Therefore the condition given by Eq.~(\ref{pm15}) imposes severe restrictions
on the form of function $f$. Assuming $\b b=\b r$,  we find as the necessary
condition
\begin{equation}\label{pm16}
f(\b p,\b r,t)=f(\b q,0,t)\,,
\end{equation}
where $\b q$ is defined in Eq.~(\ref{pb}). In the flat space $\b q$ does
not change  under shift, therefore any function of the form of Eq.~(\ref{pm16}) describes
a homogeneous distribution. This is not the case for  $\k=\pm 1$. In the curved
space the homogeneous distribution is also isotropic, i.e. $f$ depends only
on $|\b q|$ (see Appendix~\ref{ApB}). Then in the case of homogeneous distribution
one can write
\begin{equation}\label{pm18}
f(\b p,\b r,t)=g(q,t)\,.
\end{equation}
We assume further that at $\k=0$ the distribution function is also isotropic.

The substitution of Eq.~(\ref{pm18}) into Eq.~(\ref{pm14a}) 
cancels  the last two terms that results in the simple equation
\begin{equation}\label{pm19}
\left(\frac{\d g}{\d t}\right)_{\! q}=0\,.
\end{equation}
Here for clarity we introduce the notation used in thermodynamics to stress
that the time derivative is taken at constant $q$. Further we use this notation
as well. The result means that for noninteracting particles, in the case of
homogeneous and isotropic distribution,  the distribution function \emph{in
phase space} does not depend on time. It should be noted that in the case
under consideration Eq.~(\ref{pm19}) retains its form also in the presence
of arbitrary magnetic field if one abstract from energy losses due to
synchrotron radiation.

Let us derive the equation for the distribution function $F$ defined in Eq.~(\ref{pm20}).
If distribution is homogeneous, the function does not depend on $\b n$ and
$\b r$ and one can assume\footnote{It should be noted that in the case of
homogeneous distribution the collision integrals are written as in  
the flat space since $F(E,t)=F'(E,t)$.}
\begin{equation}
 F\equiv F(E,t)=4\pi PE\,g(q,t)\,.
\end{equation}
In the new variables $(E,t)$ the derivative $(\d g/\d t)_q$ is written in
the form
\begin{equation}
 \left(\frac{\d g}{\d t}\right)_{q}=\left(\frac{\d g}{\d t}\right)_E-
\frac{H P^2}{E}\left(\frac{\d g}{\d E}\right)_t\,,
\end{equation}
where $H=\dot a/a$ is the Hubble constant. Expressing  $g$ through $F$, we
find 
\begin{equation}\label{pm22a}
\left(\frac{\d g}{\d t}\right)_{q}=\frac{1}{4\pi PE}\,\hat K F(E,t)\,,
\end{equation}
where $\hat K$ is an operator defined as:
\begin{equation}\label{pm22}
 \hat KF(E,t)\equiv \left(\frac{\d F}{\d t}\right)_E-H\left(\frac{\d}{\d
 E}
\left(\frac{P^2}{E}\,F\right)_t -3F\right).
\end{equation}

The general solution of Eq.~(\ref{pm19}) is 
\begin{equation}\label{pm25}
g(q,t)=\Phi(q)\,,
\end{equation}
where $\Phi$ is an arbitrary function of one argument. Then the function
$F$ is
\begin{equation}\label{pm26}
F(E,t)=4\pi PE\,\Phi\big(a(t)\,P\big)\,.
\end{equation}
This expression defines the evolution of the distribution function for noninteracting
particles;  if the function $F(E,t')$ is known at the moment $t=t'$,  Eq.~(\ref{pm26})
allows us to find $F$ at subsequent (previous) moments of time.
One can write  $F(E,t')$ in an explicit form:
\begin{equation}\label{pm27}
F(E,t)=\frac{a(t')}{a(t)} \frac{E}{{\cal E}(E,t,t')}\,
F\!\left({\cal E}(E,t,t'),t'\right),
\end{equation}
where
\begin{equation}\label{pm27a}
{\cal E}(E,t,t')=\frac{a(t)}{a(t')}
\sqrt{E^2-m^2\left(1-\frac{a^2(t')}{a^2(t)} \right)}\,.
\end{equation}
The quantity ${\cal E}(E,t,t')$ has a simple physical meaning;  it is the energy
which particle should have at the moment $t'$ in order to have 
the energy $E$ at the moment $t$. Therefore the following relation takes place 
\begin{equation}\label{pm35}
{\cal E}\big({\cal E}(E,t,t''),t'', t'\big)={\cal E}(E,t,t')\,,
\end{equation}
For  massless particles this  equation 
is simplified:
\begin{equation}\label{pm28}
F(E,t)=\left(\frac{a(t')}{a(t)}\right)^{\!\!2} F\!\left(Ea(t)/a(t'),t'\right).
\end{equation}

The particle number density is
\begin{equation}\label{pm29}
N(t)=\int_m^{\infty}\! F(E,t)\,dE\,.
\end{equation}
Writing $F$ in the form  of Eq.~(\ref{pm27}) and introducing new variable of integration
$E'={\cal E}(E,t,t') $, we find
\begin{equation}\label{pm29a}
 N(t)=\left(\frac{a(t')}{a(t)}\right)^{\!\!3} \int_m^{\infty}\! F(E',t')\,dE' \, ,
\end{equation}
and thus
\begin{equation}\label{pm30}
N(t)\,a^3(t)=N(t')\,a^3(t')={\rm const}\,.
\end{equation}

In the presence of sources the equation for $g$ can be written in the
form
\begin{equation}\label{pm21}
\left(\frac{\d g}{\d t}\right)_{\!q}=s(q,t)\,.
\end{equation}
For  the case under consideration (homogeneous and isotropic space) the source
function $s(q,t)$ depends only on $q$ and $t$ as does the distribution function.
Assuming that the source is activated at the moment $t_i$ and that $g(q,t_i)=0$, 
we obtain
\begin{equation}\label{pm31}
g(q,t)=\int_{t_i}^t \! s(q,t')\,dt'\,,\qquad t\ge t_i\,.
\end{equation}
The equation for $F$ in the presence of the sources has the form 
\begin{equation}\label{pm32}
\hat KF(E,t)=S(E,t),\quad S(E,t)=4\pi PE\,s(q,t)\,.
\end{equation}
%где $S(E,t)=4\pi PE\,s(q,t)$.
The solution of this equation is
\begin{equation}\label{pm33}
F(E,t)=E\int_{t_i}^t \frac{a(t')}{a(t)}\, \frac{S\big({\cal E}(E,t,t'),t'\big)}
{{\cal E}(E,t,t')}\,dt'\,.
\end{equation}
Eq.~(\ref{pm33}) can be considered as Eq.~(\ref{pm31}) written in different
notations. Using Eq.~(\ref{pm35}) it is easy to verify that if the source
is active over a finite time $t_i<t<t_f$, then,  after the source is switched off 
($t>t_f$),  Eqs.~(\ref{pm33}) and (\ref{pm27}) are equivalent.

Let us consider the equation
\begin{equation}\label{pm36}
\left(\frac{\d g}{\d t}\right)_{q}=-\lambda(q,t)\,g(q,t)\,,
\end{equation}
which describes absorbtion or decay of particles. Its solution is 
\begin{equation}\label{pm37}
g(q,t)=g(q,t_i)\,e^{-\tau}\,,
\end{equation}
where the optical depth is
\begin{equation}\label{pm38}
\tau=\int_{t_i}^t \!\lambda(q,t')\,dt'\,.
\end{equation}
Denoting
\begin{equation}\label{pm39}
\lambda(q,t')=\mu(E',t') \equiv\mu\big(\sqrt{q^2/a^2(t')+m^2},t'\big)\,,
\end{equation}
and switching from variables $(q,t)$ to $(E,t)$, we obtain
\begin{equation}\label{pm40}
\tau=\int_{t_i}^t \!\mu\!\left({\cal E}(E,t,t'),t'\right) dt'\,.
\end{equation}
In the presence of absorbtion, the function  $F$ satisfies the equation 
\begin{equation}\label{pm41}
 \hat KF(E,t)=-\mu(E,t)\,F(E,t)\,,
\end{equation}
which has the solution
\begin{equation}\label{pm42}
F(E,t)=\frac{a(t_i)}{a(t)} \frac{E}{{\cal E}(E,t,t_i)}\,
F\!\left({\cal E}(E,t,t_i),t_i\right)e^{-\tau} \, .
\end{equation}

In a more general case, $g$ obey the equation
\begin{equation}\label{pm43}
\left(\frac{\d g}{\d t}\right)_{q}=s(q,t)-\lambda(q,t)\,g(q,t) \, ,
\end{equation}
and at $t<t_i$ the functions $g$ and $s$ equal to zero. Then we have
\begin{equation}\label{pm44}
g(q,t)=\int_{t_i}^t\! dt'\, s(q,t')\exp\!\left(-\int_{t'}^t\!
 \lambda(q,t'')\,dt''\right).
\end{equation}
The corresponding equation for the function $F$ 
\begin{equation}\label{pm45}
 \hat KF(E,t)=S(E,t)-\mu(E,t)\,F(E,t) \, ,
\end{equation}
has the following solution
\begin{multline}\label{pm46}
F(E,t)=E\int_{t_i}^t dt'\, \frac{a(t')}{a(t)}\, \frac{S\big({\cal
E}(E,t,t'),t'\big)} {{\cal E}(E,t,t')} \\
\times\exp\!\left(-\int_{t'}^t \! \mu\!
\left({\cal E}(E,t,t''),t''\right)dt''\right).
\end{multline}
This expression can be derived from Eq.~(\ref{pm44}) using definitions  in 
Eqs.~(\ref{pm23}) and (\ref{pm32}),  and switching from variables 
$(q,t)$ to $(E,t)$. One can
move from the integration over the time $dt$ in Eq.~(\ref{pm46}) to the integration
over the redshift $dz$ (Appendix~\ref{time}). 

It should be noted that in the case of homogeneous distribution the kinetic
equations for the metrics with $\k=0$, $+1$ and $-1$ are written identically.
However, it does not mean that the curvature does not effect on kinetics. 
In fact it does, because the behaviour of $a(t)$, which enters into the equations,
depends on $\k$.

\subsection{Energy losses}
In the case of presence of a source of particles and 
continuous energy losses the equation
for distribution function has the following form 
\begin{equation}\label{pm47}
\hat K F(E,t)-\frac{\d}{\d E}\left(b(E,t)\,F\right) =S(E,t)\,.
\end{equation}
To solve Eq.~(\ref{pm47}) let us first find the Green function $G(E,E_0,t,t_0)$
which satisfies (by definition)  the equation
\begin{equation}\label{pm48}
\hat K G-\frac{\d}{\d E}\left(b(E,t)\,G\right)=\delta(E-E_0)\,\delta(t-t_0)
\end{equation}
and the condition: $G\big|_{t<t_0}=0$. From Eq.~(\ref{pm48}) it follows that
\begin{equation}\label{pm48a}
G(E,E_0,t_0+0,t_0)=\delta(E-E_0)\,. 
\end{equation}
The Green function is the distribution function for the case when at the
moment $t_0$ one particle with energy $E_0$ is injected.

The solution is sought in the form
\begin{equation}\label{pm49}
G(E,E_0,t,t_0)=u(t,t_0)\,\delta(E-{\cal E}(E_0,t,t_0))\,\theta(t-t_0)\,,
\end{equation}
where $\cal E$ is defined  in Eq.~(\ref{pm27a}). From Eq.~(\ref{pm48a}) it
follows that the functions $u$ and $\cal E$ satisfy the initial conditions:
\begin{equation}\label{pm50}
u(t_0,t_0)=1\,,\qquad {\cal E}(E_0,t_0,t_0)=E_0\,.
\end{equation}
After substitution of Eq.~(\ref{pm49}) into Eq.~(\ref{pm48}),  it is helpful
to represent the product $b(E,t)\,G$ as $b({\cal E}(E_0,t,t_0),t)\,G$.
% and remove the factor $b({\cal E}(E_0,t,t_0),t)$ from the  
%derivative sign $\d/\d E$. 
%The same procedure should be done in the first term. 
Consequently we find  the following system of ordinary differential equations
\begin{equation}\label{pm51}
\dot u+3Hu=0\,,
\end{equation}
\begin{equation}\label{pm52}
\dot{\cal E}+\Psi({\cal E},t)=0\,,
\end{equation}
where
\begin{equation}\label{pm53}
\Psi({\cal E},t)=H({\cal E}-m^2/{\cal E})+b({\cal E},t)\,.
\end{equation}
Eq.~(\ref{pm51}) gives:
\begin{equation}\label{pm54}
u(t,t_0)=\left(\frac{a(t_0)}{a(t)}\right)^{\!3}\,.
\end{equation}
In the general case,  Eq.~(\ref{pm52})  can be  solved numerically. 

The distribution function is expressed via Green function in the following
way:
\begin{multline}\label{pm55}
F(E,t)=\int_0^t\! dt_0\,u(t,t_0)\\ \times
\int_m^{\infty}\!\!dE_0 \,\delta(E-{\cal E}(E_0,t,t_0))\,S(E_0,t_0)\,.
\end{multline}
The integration over $dE_0$ gives
\begin{equation}\label{pm56}
F(E,t)=\int_0^t\! dt_0\,u(t,t_0)\,S(E_0,t_0)\,(\d {\cal E}/\d E_0)^{-1}\,,
\end{equation}
where the solution of the equation 
\begin{equation}\label{pm56a}
{\cal E}(E_0,t,t_0)=E\,
\end{equation}
should be substituted into integrand instead of $E_0$.
%так что $E_0=E_0(E,t,t_0)$.

It is convenient to perform calculations in the following way. Let us denote
by $U(E,t_0,t)$ the solution of the equation
\begin{equation}\label{apm52}
\frac{\d U}{\d t_0}+\Psi(U,t_0)=0\,,
\end{equation}
which satisfy
\begin{equation}\label{apm53}
U(E,tе,tе)=Eе\,.
\end{equation}
In this notations we have ${\cal E}(E_0,t,t_0)=U(E_0,t,t_0)$ and the solution
of Eq.~(\ref{pm56a}) can be written in the form $E_0=U(E,t_0,t)$. For calculation
of the last factor,  we note that  
\begin{equation}\label{apm54}
{\cal D}(E,t_0,t)\equiv\frac{\d E_0}{\d E}=\frac{\d}{\d E}U(E,t_0,t)\,.
\end{equation}
This function obey the linear differential equation
\begin{equation}\label{pm57}
\frac{\d}{\d t_0}{\cal D}+\Psi_1{\cal D}=0\,,
\end{equation}
where the function 
\begin{equation}\label{pm57a}
 \Psi_1\equiv\Psi_1(U(E,t_0,t),t_0)=
 \left.\frac{\d}{\d E'}\Psi(E',t_0)\right|_{E'=U(E,t_0,t)}\,
\end{equation}
can be obtained by differentiation of Eq.~(\ref{apm52}) and after some obvious
redesignations. The initial condition has the form ${\cal D}(E,t,t)=1$, therefore
\begin{equation}\label{pm58}
{\cal D}(E,t_0,t)=\exp\!\left(\int_{t_0}^t\Psi_1(U(E,t'_0,t),t'_0)
\,dt'_0\right).
\end{equation}
If $\Psi$ does not depend explicitly on time, the equation reduces to
\begin{equation}\label{pm59}
{\cal D}(E,t_0,t)=\frac{\Psi(E)}{\Psi(E_0)}\,,
\end{equation}
but in the general case it is necessary to use Eq.~(\ref{pm58}). Thus, in
the case of energy losses, the distribution function is
\begin{equation}\label{pm60}
F(E,t)=\int_0^t\! u(t,t_0)\,S(U(E,t_0,t),t_0)
{\cal D}(E,t_0,t)\,dt_0\,\,.
\end{equation}
For derivation of  ${\cal D}$ it is easier  if one  uses  Eq.~(\ref{pm57}) with the initial
condition ${\cal D}(E,t,t)=1$ instead of the integral representation 
of Eq.~(\ref{pm58}).
Let us define the function
\begin{equation}\label{pm61}
F(E,t_0,t)=\int_{t_0}^{t}\! u(t,t'_0)\,S(U(E,t'_0,t),t_0)
{\cal D}(E,t'_0,t)\,dt'_0\,\,.
\end{equation}
It is obvious that $F(E,t)=F(E,0,t)$. The function $F(E,t_0,t)$ can be found
solving the differential equation
\begin{equation}\label{pm62}
\frac{\d}{\d t_0}F(E,t_0,t)+u(t,t_0)\,S(U(E,t_0,t),t_0){\cal D}(E,t_0,t)=0
\end{equation}
with the initial condition $F(E,t_0=t,t)=0$. Thus, the derivation of $F(E,t)$
requires a solution  of the system of three ordinary differential equation of 
first order, namely  Eqs.~(\ref{apm52}), (\ref{pm57}) and (\ref{pm62}), starting from
the point $t_0=t$, where the values of the functions are known, to the point
$t_0=0$.

\subsection{Spherically symmetric distribution\label{sphere}}

In the case of spherical symmetry,  the distribution function can be considered
as a function of the following arguments
\begin{equation}\label{sph1}
 f=f(q,r,\mu,t)\,,
\end{equation}
where $\mu=(\b p\b r)/(|\b p||\b r|)=\cos\theta$. For this  function Eq.~(\ref{pm14a})
becomes
\begin{multline}\label{sph2}
 \frac{\d f}{\d t}+\frac{1}{a(t)\sqrt{(1+m^2a^2(t)/q^2)(1-kr^2\mu^2)}}\\
\times\left(\mu(1-kr^2) \frac{\d f}{\d r}
 +\frac{1-\mu^2}{r}\frac{\d f}{\d\mu} \right)=0.\quad
\end{multline}
The equation does not contain the derivative $\d f/\d q$, i.e. $q$ enters
in this equation as a parameter. Therefore it is helpful to introduce the
new variable instead of time
\begin{equation}\label{sph3}
 \eta=\int_{t_i}^t \frac{dt'}{a(t')\sqrt{1+m^2a^2(t')/q^2}}
\end{equation}
and consider $f$  as a function of the arguments $(q,r,\mu,\eta)$. Then it
brings us to the equation  
\begin{equation}\label{sph4}
 \frac{\d f}{\d\eta}+\frac{1}{\sqrt{1-kr^2\mu^2}}\left(\mu(1-kr^2)
\frac{\d f}{\d r} +\frac{1-\mu^2}{r}\frac{\d f}{\d\mu} \right)\!=\!0,
\end{equation}
which admits a solution in the general case. The general solution of the
equation determined by the method of characteristics has the form
\begin{equation}\label{sph5}
 f(q,r,\mu,\eta)=\Phi(q,X,Y)\,,
\end{equation}
where
\begin{equation}\label{sph6}
 X=\frac{r^2(1-\mu^2)}{1-\k r^2}\,,
\end{equation}
\begin{multline}\label{sph7}
Y=\frac1{\sqrt{-\k}}\ln\left(\sqrt{1-\k r^2\mu^2}+r\mu\sqrt{-\k}\right)-\eta\\
=\frac1{\sqrt{-\k}}\,{\rm arsinh}\left(r\mu\sqrt{-\k}\right)-\eta\,,
\end{multline}
$\Phi$ is an arbitrary function of three arguments\footnote{This solution
is surely a special case of Eq.~(\ref{hj7}), but it is easier to obtain it
by solving directly Eq.~(\ref{sph4}).}. In the case of $\k=1$,  we have $\sqrt{-\k}=i$,
and it is convenient to present Eq.~(\ref{sph7}) in the form
\begin{gather}\label{sph8}
 Y=\arcsin(r\mu)-\eta\, .
\end{gather}
For $\k=0$ we have 
\begin{equation}\label{sph9}
 Y=r\mu-\eta\,.
\end{equation}

The obtained results are quite convenient to use at least in two cases.

\subsubsection{The boundary problem}

Let us assume that the distribution function is known at a certain point $r=r_*$
(on the surface of a ``star''):
\begin{equation}\label{sph10}
 f(q,r_*,\mu,\eta)=f_*(q,\mu,\eta)\,.
\end{equation}
Then Eq.~(\ref{sph5}) allows us to derive the distribution function in all space.
For instance, in the case $\k=-1$ we get
\begin{equation}\label{sph11}
 f(q,r,\mu,\eta)=f_*(q,\t\mu,\t\eta)\,,
\end{equation}
where
\begin{equation}\label{sph12}
 \t\mu=\frac{r\mu}{r_*^{}\,\sqrt{1+r^2}}\sqrt{1+r_*^2-\frac{r^2-r_*^2}{r^2\mu^2
} } \,,
\end{equation}
\begin{equation}\label{sph13}
 \t\eta=\eta+\ln\frac{\sqrt{1+r_*^2\t\mu^2}+r_*^{}\t\mu}
 {\sqrt{1+r^2\mu^2}+r\mu}\,.
\end{equation}

Eq.~(\ref{sph11}) defines the distribution function in the reference frame
with the origin of coordinates at the center of the ``star''. If the observer is located
at the distance $r$, then, as indicated before, it is convenient to
move to the reference frame related to  the observer. At the location point 
of  observer, the distribution function is 
\begin{equation}\label{sph14}
 f'(q,\b r'=0,\cos\theta',\eta)=f(q,r,\mu,\eta)\,,
\end{equation}
where $\mu$ ($\mu=\cos\theta$) should be represented in the form of Eq.~(\ref{pb3}).
Here, we  take into account that $q$ and $\eta$ do not change  under shift.

In the case of $\k=1$,  the following relations should be used in Eq.~(\ref{sph11}): 
\begin{equation}\label{sph12a}
 \t\mu=\frac{r\mu}{r_*^{}\,\sqrt{1-r^2}}\sqrt{1-r_*^2-\frac{r^2-r_*^2}{r^2\mu^2
} } \,,
\end{equation}

\begin{equation}\label{sph13a}
 \t\eta=\eta+\arcsin(r_*^{}\t \mu)-\arcsin(r\mu)\,.
\end{equation}

\subsubsection{Initial value problem}

Let us assume that the distribution function is known at some moment of time
$t=t_i$ which corresponds to $\eta=0$:
\begin{equation}\label{sph15}
 f(q,r,\mu,\eta=0)=f_0(q,r,\mu)\,.
\end{equation}
Then at an arbitrary moment of time
\begin{equation}
 f(q,r,\mu,\eta)=f_0(q,r_0^{},\mu_0^{})\,.
\end{equation}
Here
\begin{equation}
 r_0^{}=\left[\frac{r^2(1-\mu^2)+(1-\k r^2)\,\zeta^2 }
{1-\k r^2\mu^2}\right]^{\!1/2}\!,\quad
 \mu_0^{}=\frac{\zeta}{r_0^{}}\,,
\end{equation}
where
\begin{equation}
\! \zeta=r\mu\,\cosh(\eta\sqrt{-\k})-
 \sqrt{\frac{1-\k r^2\mu^2}{-\k}}\, \sinh(\eta\sqrt{-\k})\,.
\end{equation}
For analysis of the angular and energy distribution it is convenient, as
before, to introduce the new reference frame choosing the origin of coordinates
at the point where the observer is located.

\section{Summary}

In this paper the mechanics and kinetics in the FLRW space-times have been
studied on the basis of the standard canonical formalism. The Cartesian coordinates
$\b r$ and the corresponding generalized momenta $\b p$ are  considered as
the points $(\b p,\b r)$  in  the phase space where the distribution function
$f(\b p,\b r,t)$ given by Eq.~(\ref{pm13}) is introduced. The form-invariance of 
equations of mechanics and kinetics relative to shift of the origin of reference frame
as well as the invariance of the distribution function $f$ have been proved. 
The transformation of the momentum under shift is described by the quite lengthy
equation, Eq.~(\ref{pm7}). But for applications it is sufficient to use
Eq.~(\ref{pb}) which defines the transformation of momentum under the shift
of the origin to the point $\b r$ where the observer is located. The collisionless
Boltzmann equation  admits general solutions for the function $f$  (see
Eq.~(\ref{hj7})).

Along with the distribution function $f$,  the ``conventional''
distribution function  $F(E,\b n,\b r,t)$ given by Eq.~(\ref{pm20}) is introduced in the
phase space. This function is more convenient for inclusion of collision integrals, 
and  defines the relationship between the functions $f$ and $F$. If the collision integral
$I$ for the Minkowski space is known, it can be found also in the FLRW space-time using the following procedure.  The origin of coordinates should be shifted to the collision
point where $I$ can be written as in the flat space-time. After that the
collision integral should be transformed to the initial reference frame using
the formulas  obtained in this paper.

The equations are considerably simplified in the case of homogeneous and
isotropic distribution. For this case the analytical
solution of the kinetic equation with the source and absorption processes
is given in Section~\ref{uniform}. In the case of energy losses the equation
can be no longer solved analytically and determination of $F$ comes to solving
the system of three ordinary differential equations.

The results of Section \ref{sphere} can be quite useful for analysis of angular
and energy  distribution of  particles  from sources located at cosmological
distances. The distribution function represented by Eq.~(\ref{sph5}) describes the solution
in the reference frame related to  the source. This function can be easily
transformed to the reference frame associated with observer that gives the
spectrum and angular  distribution in the observation point.
%qqq
% \section{Acknowledgments}

\appendix

\section{Relativistic Boltzmann equation in the general form \label{ApA}}
Below a simple derivation of the collisionless Boltzmann
equation for the space with an arbitrary metric  $g_{\alpha\beta}(x)$ is
presented. It is shown that the equation can be written in the form of 
Eq.~(\ref{pm14a}) with a relevant Hamiltonian function.
Let $x^{(s)i}(t)_{}$ ($i\!=\!1,2,3$) are the coordinates of the particle with
number $s$ at the moment $t$, and let $p^{(s)}_i(t)$ are the  covariant components of
its momentum. It is convenient to consider the single-particle distribution
function \cite{Debbasch1,Debbasch2} as a function of the following independent
variables: covariant momentum  $\b p=(p_1,p_2,p_3)$, coordinates $\b r=(x^1,x^2,x^3)$
and time. Let us also introduce the zero components
\begin{multline}\label{ap0}
x^0=t\,, \qquad p_0^{}\equiv p_0^{}(\b p,\b r,t)=\frac1{g^{00}}\Big(-g^{0i}p_i+
\\ \sqrt{(g^{0i}p_i)^2- g^{00}(g^{ik}p_ip_k-m^2)} \,\Big).\qquad{}
\end{multline}
Here  $p_0^{}$ is the solution of the quadratic equation 
$g^{\alpha\beta}p_\alpha p_\beta =m^2$, and the solution should be chosen
from two possible ones,  so that $p^0=g^{0\alpha}p_\alpha>0$.

The microscopic single-particle distribution function is defined from the  following
relations: 
\begin{gather}\label{ap1}
 f(\b p,\b r,t)=\sum_s f^{(s)}(\b p,\b r,t)\,,
 \end{gather}
\begin{gather}\label{ap2}
 f^{(s)}(\b p,\b r,t)=\delta(\b r-\b r^{(s)}(t))\,\delta(\b p-\b p^{(s)}(t))\,,
\end{gather}
where three-dimensional $\delta$-functions are equal to product of three
one-dimensional ones. For derivation of the equation let us differentiate
$f^{(s)}$ with respect to time, 
\begin{equation}\label{ap3}
 \frac{\d f^{(s)}}{dt}=-\frac{\d}{\d x^i}(\dot{x}^{(s)i}f^{(s)})-
 \frac{\d}{\d p_i}(\dot{p}^{(s)}_if^{(s)})\,,
\end{equation}
where for  the convenience of further transformations the quantities $\dot{x}^{(s)i}$
and $\dot{p}^{(s)}_i$ are inserted under the sign of partial derivatives.
From the definition of momentum it is follows that $\dot{x}^{(s)i}=p^{(s)i}/p^{(s)0}$,
where $p^{(s)\alpha}=g^{\alpha\beta} p^{(s)}_\beta$ are contravariant components
of the 4-momentum. Taking into account
that $\delta$-functions entering into $f^{(s)}$ allow to replace $\b r^{(s)}(t)$
and $\b p^{(s)}(t)$ by $\b r$ and $\b p$, in Eq.~(\ref{ap3})  $p^i/p^0$ can
be written instead of  $\dot{x}^{(s)i}$  (here the following relation 
is used: $\delta(x-x_0)\,f(x)=\delta(x-x_0)\,f(x_0)$).

The same operation can be done in the last term of Eq.~(\ref{ap3}): 
\begin{equation}\label{ap4}
\dot{p}_i^{(s)} =\frac{1}{p^{(s)0}}\,\Gamma_{\alpha,\beta i}
p^{(s)\alpha}p^{(s)\beta} = \frac1{2p^{(s)0}}\, \frac{\d g_{\alpha\beta}}
{\d x^i} p^{(s)\alpha}p^{(s)\beta},
\end{equation}
where $\Gamma_{\dots}$ are the Christoffel symbols; the 
values of the functions
should be taken at the point  $(\b p^{(s)}(t),\b r^{(s)}(t))$. The presence of
$\delta$-functions allows the following replacement   
\begin{equation}\label{ap5}
\dot{p}_i^{(s)}\to \frac1{2p^{0}}\,\frac{\d g_{\alpha\beta}}
{\d x^i}\,p^{\alpha} p^{\beta}\,,
\end{equation}
where all functions are evaluated at the point $(\b p,\b r)$. Thus it brings
us to the equation
\begin{equation}\label{ap6}
 \frac{\d f^{(s)}}{dt}+\frac{\d}{\d x^i}\Big(\frac{p^i}{p^0}\,f^{(s)}\Big)+
 \frac{\d}{\d p_i}\Big(\frac1{2p^{0}}\,
 \frac{\d g_{\alpha\beta}}{\d x^i}\,p^{\alpha} p^{\beta}f^{(s)}\Big)=0.
\end{equation}

By differentiating the equation
$g^{\alpha\beta}p_\alpha p_\beta =m^2$ with respect to $p_i$, we find 
\begin{equation}\label{ap7}
 g^{\alpha\beta}p_\alpha\,\frac{\d p_\beta}{\d p_i} =
 p^0\,\frac{\d p_0^{}}{\d p_i}+p^i=0\,.
\end{equation}
By differentiating the equation $g_{\alpha\beta}p^\alpha p^\beta =m^2$  with
respect to $x^i$, we get
\begin{equation}\label{ap8}
 \frac12\,\frac{\d g_{\alpha\beta}}{\d x^i} p^\alpha p^\beta =
 -p_\alpha\frac{\d p^\alpha}{\d x^i}= p^\alpha\frac{\d p_\alpha}{\d x^i}
 =p^0\frac{\d p_0^{}}{\d x^i}\,,
 \end{equation}
where it is taken into account that $\d p_k/\d x^i=0$. Using Eq.~(\ref{ap7}) and 
(\ref{ap8}),  Eq. (\ref{ap6}) can be written in the form
\begin{equation}\label{ap9}
 \frac{\d f^{(s)}}{\d t}-\frac{\d}{\d x^i}\Big(\frac{\d p_0^{}}{\d p_i}
\,f^{(s)}\Big)+\frac{\d}{\d p_i}\Big(\frac{\d p_0^{}}{\d x^i}f^{(s)}\Big)=0.
\end{equation}
It can be seen that the terms, which do not contain derivatives of  $f^{(s)}$,
are canceled.

Summing over $s$, we obtain the equation for the function given by Eq.~(\ref{ap1}):
  \begin{equation}\label{ap10}
  \left(\frac{\d}{\d t}-\frac{\d p_0^{}}{\d p_i}\,\frac{\d}{\d x^i}+
  \frac{\d p_0^{}}{\d x^i}\,\frac{\d}{\d p_i}\right)\! f=0\,.
  \end{equation}
This equation can be also written in the form
  \begin{equation}\label{ap11}
 \left(p^{\alpha}\frac{\d}{\d x^\alpha}+\frac12 \frac{\d g_{\alpha\beta}}
 {\d x^i}\,p^{\alpha} p^{\beta}\frac{\d}{\d p_i}\right)\! f=0\,.
 \end{equation}
The macroscopic (averaged over the ensemble) distribution function satisfies
the same equations.

Note that the second and third terms enter into Eq.~(\ref{pm14a}) and Eq.~(\ref{ap10})
with opposite signs. The reason  is the following.  Throughout the  paper
we use the generalized momentum $\d L/\d v^i$,  while $p_i$ in this Appendix is the
covariant components of momentum. One can show that for the Lagrangian 
of the general form
\begin{equation}\label{ap12}
 L=-m\sqrt{g_{00}^{}+2g_{0i}^{}v^i+g_{ik}^{}v^i v^k}
\end{equation}
the generalized momentum differs from  $p_i$ by sign: $\d L/\d v^i=-p_i$.
Therefore the momenta used in the main part of the paper and in this appendix have opposite signs.
The energy $E=v^i\d L/\d v^i-L$ coincides with $p_0^{}$, and consequently,
if one replaces  $p_i\to -p_i$ in Eq.~(\ref{ap0}), we will  get  the Hamilton
function expressed through generalized momenta and coordinates.
In this case the Hamilton equations and the Boltzmann equation have 
standard forms. 

\section{Isotropy of homogeneous distribution\label{ApB}}

It is easy to show that in the case of $\k=\pm 1$ the direction of the vector
$\b q$ changes under shift. It suffices to consider the infinitesimal shift
$\delta\b b$. Let us denote by $\delta\b q$ the change of  $\b q$ under the
shift to $\delta\b b$. Since $\b q^2$ is invariant relative to shift, $\delta\b
q^2=2(\b q\,\delta\b q)=0$, i.e. the vectors  $\b q$ and $\delta\b q$ are
orthogonal. Therefore $\delta\b q$ can be represented as 
\begin{equation}\label{app1}
 \delta\b q=(\delta\b\phi\times\b q)\,,
\end{equation}
where $\delta\b\phi$ depends linearly on $\delta\b b$ and can be considered
as an arbitrary vector. The requirement of invariance of $f$ relative to
the shift gives 
\begin{equation}\label{app2}
 \delta f=\delta\b q\,\frac{\d f}{\d\b q}=
 (\delta\b\phi\times\b q)\, \frac{\d f}{\d\b q}=
 \delta\b\phi\,\Big(\b q\times \frac{\d f}{\d\b q}\Big) =0\,,
\end{equation}
that is equivalent, due to the arbitrariness of $\delta\b\phi$, to the equality
\begin{equation}\label{app3}
\b q\times\frac{\d f}{\d\b q}=0\,.
\end{equation}
This implies that the vector $\b q$ and $\d f/\d\b q$ are parallel,  and
\begin{equation}\label{app4}
 \frac{\d f}{\d\b q}=\b q\,A(\b q,t)\,.
\end{equation}
Writing this equality in spherical coordinates, one can ascertain that  
$f$ does not depend on angular variables, i.e. it is a function of only $|\b
q|$. Thus, we conclude that the isotropy of the distribution follows from
its homogeneity.

\section{Connection between time and redshift \label{time}}

For the flat $\Lambda$CDM model, the Friedmann equation has the following 
form (see, for example, Ref.~\cite{Carroll})
\begin{equation}\label{rs1}
\left(\frac{\dot a}{a}\right)^{\!2}=H_0^2\left[\Omega_m
\left(\frac{a(t_0)}{a(t)}\right)^{\!3}+\Omega_\Lambda \right],
\end{equation}
where $a(t_0)$ is the value of the scale factor $a$ at the present epoch.
This equation can be solved analytically. Let us introduce the function
$\beta(t)=\big(a(t)/a(t_0)\big)^{3/2}$. For $\beta$ we have the following  equation
\begin{equation}\label{rs2}
\dot\beta=\frac32\,H_0\sqrt{\Omega_\Lambda\beta^2+\Omega_m}\,,
\end{equation}
from where, taking into account that  $\beta(0)=0$, we find
\begin{equation}\label{rs3}
\frac1{\sqrt{\Omega_\Lambda}}\ln\!\left(\frac{\beta\sqrt{\Omega_\Lambda
}+\sqrt{\beta^2\Omega_\Lambda +\Omega_m}}{\sqrt{\Omega_m}}\right)
=\frac32\,H_0t\,.
\end{equation}
Solving Eq.~(\ref{rs3}) for $\beta$, we obtain
\begin{equation}\label{rs4}
\frac{a(t)}{a(t_0)} =\left(\frac{\Omega_m}{\Omega_\Lambda}\right)^{1/3}
\left[\sinh\!\left(\frac32\sqrt{\Omega_\Lambda}\,H_0t\right)\right]^{2/3}\,.
\end{equation}
This expression is also derived in \cite{Carroll} in a different way (see
Eq.~(29.131)).  According to WMAP \cite{WMAP11s}, 
the parameters in Eq.~(\ref{rs4}) have the following values: $H_0=71 \ {\rm km\; s^{-1} \,Mpc^{-1}}$,
$\Omega_m=\Omega_b+\Omega_c=0.27$, $\Omega_\Lambda=0.73$. In the 
model of flat Universe  the scale factor $a$ is determined with an accuracy of 
an arbitrary factor, therefore it is convenient to adopt  $a(t_0)=1$,
that is equivalent to the redefining of
the comoving coordinates $\b r$.

The redshift $z$ and $a(t)$ are connected as 
\begin{equation}\label{rs5}
1+z=a(t_0)/a(t)\,.
\end{equation}
If $a(t)$ is monotonically increasing, the connection between $z$ and $t$
is unique, and one can move from the integration over $dt$ to the integration
over $dz$. In the case of flat Universe  we have
\begin{equation}\label{rs6}
 dt=-\frac{dz}{H_0(1+z)\sqrt{\Omega_m(1+z)^3+\Omega_\Lambda}}\,.
\end{equation}

Naturally, the results obtained by integrating over time with  
Eq.~(\ref{rs4}) coincide with ones obtained by integrating over $dz$.

\section{Superluminal recession velocity \label{ApD}}

In the expanding Universe distant objects have superluminal recession velocities.
This issue is elucidated quite comprehensively in \cite{Davis}. Here we want to 
emphasize that superluminal recession velocities always occur in the expanding space with
$\k=0$ and $\k=-1$, and to call attention to some features of photon propagation from
sources with superluminal recession velocities. Let us assume that the observer
registers  a photon at  $t=t_o$. Then,  at the moment $t<t_o$ the proper
distance from the photon to the observer equals to 
%(conventional units)
\begin{equation}\label{bp1}
 R_p(t)=a(t)\int_t^{t_o}\!\frac{c\,dt'}{a(t')}\,.
\end{equation}
If $a={\rm const}$, we would have $R_p(t)=c(t_o-t)$, i.e. $R_p(t)$ decreases
linearly. If $a(t)$ increases with time, and $a(0)=0$, then the propagation of
the photon is qualitatively different. In this case, as it follows from Eq.~(\ref{bp1}),
$R_p(0)=0$, $R_p(t_o)=0$, and in the range $0<t<t_o$ the function $R_p(t)>0$.

\begin{figure}
 \begin{center}
 \includegraphics[width=0.32\textwidth,angle=-90]{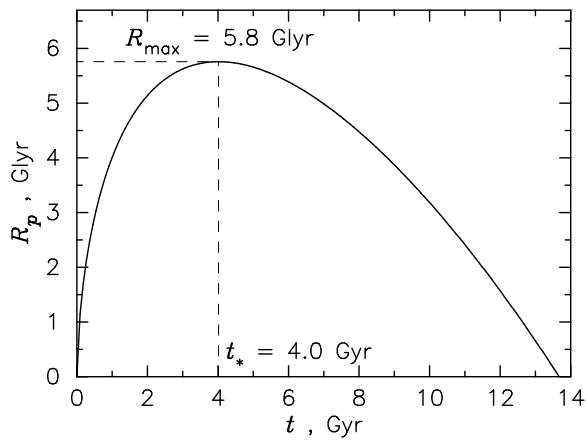}
 \caption{The dependence of proper distance on time \label{r_t}}
 \end{center}
 \end{figure}

Fig.~\ref{r_t} shows the time dependence $R_p(t)$. The calculations are performed
for the values of parameters  used in Appendix~\ref{time}, however, it is
clear that the qualitative behavior of the curve is defined only by the condition
$a(0)=~0$.

Let us call by proper velocity, $v_p^{}$, the time derivative of the proper
distance:
\begin{equation}\label{bp2}
 v_p^{}(t)\equiv dR_p/dt=v_r(t)-c\,,
\end{equation}
where $v_r(t)=(\dot a(t)/a(t))\,R_p(t)=H(t)\,R_p(t)$ is the recession velocity,
$H(t)$ is the Hubble constant. At the point of maximum $t=t_*$ one has
$v_p(t_*)=0$, therefore $v_r(t_*)=c$.  In the range $0<t<t_*$ the quantity
$v_p>0$. It means that the recession velocity is greater than $c$. The moment
$t_*$ corresponds to the redshift $z_*=1.64$, i.e. all sources with $z>1.64$
move away from us with velocities greater than the speed of light. 
Photons emitted by this
sources in the direction to the observer initially move away, reaching the
maximum proper distance $R_{\max}$, and only after that $R_p(t)$ starts
to decrease. One can show  that the source has a minimal angular size if it is located at the distance $R_{\max}\approx1.78$ Gpc that corresponds to redshift $z_*=1.64$.

In the case of the closed space ($\k=+1$),  sources with superluminal recession
velocities are also possible provided that  the condition $\dot a>c/\pi$ is fulfilled.

\end{document}